\documentclass[aps,showpacs,prl,toolkits,nofootinbib,preprint]{revtex4}

\usepackage{graphicx}
\usepackage{mathrsfs}
\usepackage{amsmath}
\usepackage{color}
\usepackage{epsfig}
\usepackage{epstopdf}
\usepackage{subfigure}

\begin{document}

\title{Searching direction dependent daily modulation in dark matter detectors}

\author{Frank T. Avignone III and Richard J Creswick}
\address{Physics Department, University of South Carolina, Columbia SC 29208, USA}
\author{Shmuel Nussinov}

\address{Tel Aviv University, Sackler School Faculty of Sciences,
Ramat Aviv, Tel Aviv 69978, Israel}

\email{nussinov@post.tau.ac.il}

\begin{abstract}

  Channeling of recoil ions in various crystals strongly depends on their
  direction. This, along with the WIMP wind due to the rotation of the
  galactic disc generates {\it daily} modulations of the channeling. Since
  channeling affect the characteristics of the WIMP induced recoil
  events these modulations may be observed in single crystal detectors and
  serve as a WIMP signature. We suggest in particular searching for such modulations 
  in the DAMA scintillation data. 

\end{abstract}

\maketitle

\subsection{I. Introduction}

  Astrophysical evidence strongly suggests non-baryonic matter constituting most of the matter's total cosmological mass density.
 Direct searches for such dark matter particles in low background, large and very sensitive underground detectors has been ongoing for several decades.
  Signatures separating signals from background can help this search via rare nuclear collisions of the Weakly Interacting Massive dark matter Particles (WIMP) in these detectors.

 One such signature, a $\sim$ 10\% annual modulations was suggested some time ago in a beautiful paper (DFS).\cite{DFS} DFS assumed isotropic WIMP velocity distribution in the halo frame.
Since the solar system is rotating with the galactic disc at a velocity $\vec{W}$ of 230 Km/sec, WIMPs impinging on earth have this net {\it wind} of velocity $\vec{W}$.  On June 2nd  a component comprising $\sim$ 1/2 of earth's rotation velocity of 30 Km/sec adds up to $W$ increasing the WIMP flux,  energy and counting rate and the opposite happens around December 2nd.
Such modulations have been seen in DAMA's Na-I scintillation experiment\cite{DAMA1}  but not in other WIMP 
searches\cite{CDM1,CDM2,CDM3,XENON,COGENT}.

 ``Channeling", namely, the tendency of atoms, ions or electrons entering a crystal to move in channels along crystal axes or Bragg planes is well known and studied. See, {\it eg.,} Refs. \cite{Lin}, \cite{And} and \cite{Hogg}.

 Channeled recoils collide less frequently and deposit more energy in electromagnetic modes. This and the enhancement of channeling at lower energies may help explain the above discrepancy.

 The possible relevance of channeling to the DAMA experiment was recently emphasized by Drobyshevski.\cite{Drob} He recalled the fact that ions entering single-crystals along crystal axes penetrate much deeper than the same ions at the same energy entering an amorphous form of the same material.  He cited the early observations of Bredov and Okuneva\cite{Bred} that 4 keV Cs$^+$ ions can penetrate $\sim 10^3$ Angstrom in Germanium crystals versus $\sim$ 44 Angstroms in amorphous Germanium. This ``channeling" effect can significantly influence the ionization signals in germanium and enhance the scintillation output in NaI(Tl) single crystals.
 Specifically, channeling can dramatically increase the ``quenching factor",
namely, the ratio of the (e.m.) energy measured from a recoiling ion in the detector to that from an incident photon of the same energy.
The quenching factor was measured in NaI(Tl) by a number of authors (see, {\it eg.,} Ref. \cite{Sim} and most recently Ref. \cite{Chag}) to be in the 19\%-26\% range.

The calculation presented by Bernabei, et al.\cite{DAMA2} suggests that for Na or I ion recoiling along a channel the quenching factor may rise to be of order unity.
If this is indeed correct, channeling which is more prevalent  at lower energies generates events with higher scintillation outputs.
Further, if WIMPS are light, $\sim$ 5 Gev, then the recoil energies, $\sim$ 1/m  in experiments utilizing heavy nuclear targets of mass m $\sim$Am$_{nucleon}$ are too low. However with the relatively light Na A $\sim$ 26 these WIMPS can be detected in DAMA via the (channeling enhanced) scintillation signal.\cite{P-Z}

In this note we focus on the strong ``directional" dependence of channeling.
 If enhanced channeling---accompanied by enhanced scintillation---is so  important as the above might suggest, then the directional dependence of  channeling might generate observable daily modulations.

 More generally the prospect of a striking temporal variation of (the e.m.~part of) the WIMP signal is interesting and may provide  yet another helpful WIMP signature.

\subsection{II. The Basic Idea and Some Kinematics}

 On the astronomical side our suggestion relies on the ``WIMP wind" velocity $\vec{W},|W|$ = 230 Km/sec. Also the earth's rotation N-S axis makes a substantial angle of $\sim 44^o$ with $\vec{W}$ and the direction of $\vec{W}$ relative to the crystalline axes changes every day.

 The other physics ingredient is the direction dependence of channeling.
Ions recoiling at angles smaller than $\theta(a)$ (or $\theta(p)$) relative to a crystal axis or to a crystal plane are ``Axially"  or ``Planarly" channeled.
These ``acceptance"  angles decrease with increasing ion velocity and both planar and axial channeling probability ($\sim \theta(p)$ and $\theta(a)^2$) fall off as $1/V \sim 1/(E_R^{1/2})$.

 For a 4 keV Iodine ion in the cubic Na-I lattice, the recipe of Appelton, et al.\cite{App} as quoted in Ref.~\cite{Drob}  yields for the main ``axial" channels along the X,Y and Z directions $\theta(a)_{100} = 6.4^o$ and for the other axes $\theta(a)_{110}=4.9^o$;  $\theta(a)_{111} = 2.8^o$.  The estimated planar channeling opening angles for the principle symmetry planes are:
$\theta(p)_{100} = 4.1^o$; $\theta(p)_{110}=3.2^o$ and $\theta(p)_{111}=2.7^o$.

 Since the 100 planes are further apart and have a larger density of Ions  the above pattern is qualitatively understood.

 The above implies that an I$^-$ Ion at a random direction  has a probability of:

\noindent $2\pi \cdot [2 \cdot (4.1^o) \times 3$ (for the three 100 planes) $ + 2 \cdot (3.1^o) \times 6$ (for the six  110 planes) $+ 2 \cdot (2.7^o) \times 4$ (for the four 111 planes)] / $(4\pi \cdot 56^o \cdot 4) \sim 20\%$

\noindent to channel---namely, to be in the union of the cones and annular patches of  all symmetry axes/planes.
(The $56^o$ in the denominator is a radian in degrees and the extra $1/4$ is due to the fact that an ion incident in any direction ``sees" only $\pi/(4 \cdot \pi) = 1/4$  of the sphere.)

 We are interested not in this overall averaged channeling probability  but in its directional dependence and the resulting temporal modulations.

 Consider Ions recoiling with a specific velocity vector $\vec{V}$ in  space. As the earth rotates $\vec{V}$ describes a cone intersecting  the unit sphere at latitude $\beta = \arccos[(V(||))/(|V|)]$ with $V_{||}$  the component of $V$ parallel to the equatorial plane.
 As $\vec{V}$ moves in and out of any one of the above cones/patches,  channeling along particular axes or planes turns on and off.
 Thus we expect not merely a 20\% time averaged  channeling  probability but rather 100\% channeling during a very specific 20\%  temperate subset of the day!

 To most easily visualize this, consider the artificial case $\beta = 0$,  namely, $\vec{V}$ lies {\it in} earth's  equatorial plane which we further take to be  the crystal X-Y plane (with the 001 axis usually pointing along the  vertical, this would happen for experiments at the north or south pole).

 Let us further assume that at 6 AM $\vec{V}$ points in the X direction.
 The above channeling acceptance angles imply that axial channeling in the  100 (+X) axis and planar channeling in the 010 (XZ) plane take place in  the 25.6 and 16.4 minutes time intervals, respectively, prior to and after 6 AM.
 After a 45$^o$ rotation or three hours later, namely, at 9 AM +/- 12.8  minutes, we have planar channeling in the 011 plane (at 45$^o$  with + X and +Y).
 At 12 PM, the 6 AM channeling-time pattern repeats, now  with the 010 Z axis  and (100) YZ plane. At 3 PM we repeat the 9 AM pattern with the -X,Y  plane, etc., so that channeling is modulated with a six-hour period.

Different computable daily modulations of channeling arise for the realistic $\beta \sim 45^o$ and other geographical latitudes (where the artifact of having continuous channeling in the XY plane as in the above is avoided). We deal later with Grand Sasso's location and the actual $\beta \sim 45^o$ angle between $\vec{W}$ and the earth's rotation axis.

 The above is reminiscent of the Bragg enhancements at certain angles (or times) of Primakoff converted solar axions in Germanium crystals used  by the Solax collaboration\cite{Sol},\cite{Cre} to improve the sensitivity of underground axion searches.

 Unfortunately, unlike the case where all photons point to the sun,  not all ions recoil in the same direction!

 The velocity $\vec{V}$ of the WIMPs is the sum of wind velocity  $\vec{W}$ and $\vec{U}$, the WIMP velocity in the halo:
\begin{equation}
  \vec{V} = \vec{W} + \vec{U}                               
\label{WIMPvel}
\end{equation}

 For all WIMPs to be incident in the direction of $\vec{W}$ we need  that (i) $\vec{U}$ be parallel to  
 $\vec{W}$ {\it and} (ii) the recoil  momentum $\vec{q}$ be parallel to $\vec{V}$---the velocity of the WIMP.

 However, the halo WIMP velocity $\vec{U}$ in the halo frame is assumed  to have an isotropic Maxwellian distribution:
\begin{equation}
  f(U) \sim U^2 \exp-(U^2/(U_0^2))                        
\label{MaxwellDist}
\end{equation}
with $U_0 \sim $ 220 Km/sec, 
and also the recoil only partially tracks the direction of the WIMP.

To find the angular distribution of the recoil ions with respect to the direction of the WIMP wind $\vec{W}$ we integrate over the above distribution subject to the kinematics of the reaction:

\noindent WIMP (momentum $\vec{p}$,mass M) + Ion(at rest,mass m) $\rightarrow$ final WIMP ($\vec{p'})$ + recoil Ion($\vec{q}$)

\noindent with
\begin{equation}
 \vec{p} = M(\vec{W} + \vec{U})
\label{vecp}
\end{equation}
Momentum conservation then fixes $\vec{p'}$ for a given recoil $\vec{q}$.

The remaining relevant constraint is that of energy conservation:
\begin{equation}
 \delta[(p^2 -(p-q)^2)/(2M) - q^2/(2m)].
 \label{EC}
\end{equation}
The uniform angular part of the $U$ integration over this $\delta$ function yields $\Theta$ functions requiring $u$ to be larger than the (absolute value of)
\begin{equation}
 q/\mu - W \cos(\theta)
\label{ThetaFunction}
\end{equation}
where $\theta$ is the angle of interest between the recoil and the WIMP wind, and $\mu = Mm/(M+m)$ the reduced mass of the WIMP-nucleus system.
Only $U$'s larger than the above can (anti-) align with  the direction of the recoil so that when added the WIND component $W \cos(\theta)$  yields the desired recoil momentum $q$.

 The remaining integration over the gaussian yields the angle-momentum distribution,
\begin{equation}
 dn/(q^2 \,  dq \, d \, \cos(\theta)) = \exp(-[q/(2\mu)- W \, \cos(\theta)]^2/(U_0^2)) \; S(q^2) \, .  
\label{AMdist}
\end{equation}


The factor $S(q^2)$ represents nuclear form factor effects roughly parameterized by $\exp(-(q^2 \cdot R^2))$  with $R \sim 1.2 \, A^{1/3}$ Fermi, the nuclear radius.
For the low recoil momenta---dominating DAMA's events, and for which significant channeling is expected---the FF effects are small and can be neglected.  Indeed it was claimed\cite{P-Z} that only relatively light $\sim$ 5 GeV WIMPS with the standard halo velocity profiles are consistent with DAMA's modulation and {\it all} other experiments (providing that channeling is indeed important and strongly enhances scintillations).  In this case,
$\mu \sim M$ and the average recoil momentum $q = 2 \mu W \sim$ 10 MeV with no significant FF effects, even for Iodine with R $\sim$ 6 Fermi. The corresponding recoil energies are $\sim$ 2 keV for Na and $\sim$ 0.5 keV for Iodine.

Using $U_0 \sim W$ since both are close to 230 km/sec, the angular distribution
\begin{equation}
 \exp(-[x-\cos(\theta)]^2    
\label{angdis}
\end{equation}
falls by, $1/e$ and $(1/e)^2 \sim$ 0.125  between $\theta=0$ and $\theta =\pi/2$ for $x=q/(2\mu \cdot W)$ = 1 and 2, respectively.

For $q \sim 0, \;  \vec{W} = \vec{U}$ and there is no forward peaking of $q$.
 The overall angular distribution obtained by integrating Eq. (\ref{AMdist}) 
 over $q$ is reasonably peaked at $\theta=0$.

\begin{figure}
\includegraphics[width=0.7\textwidth]{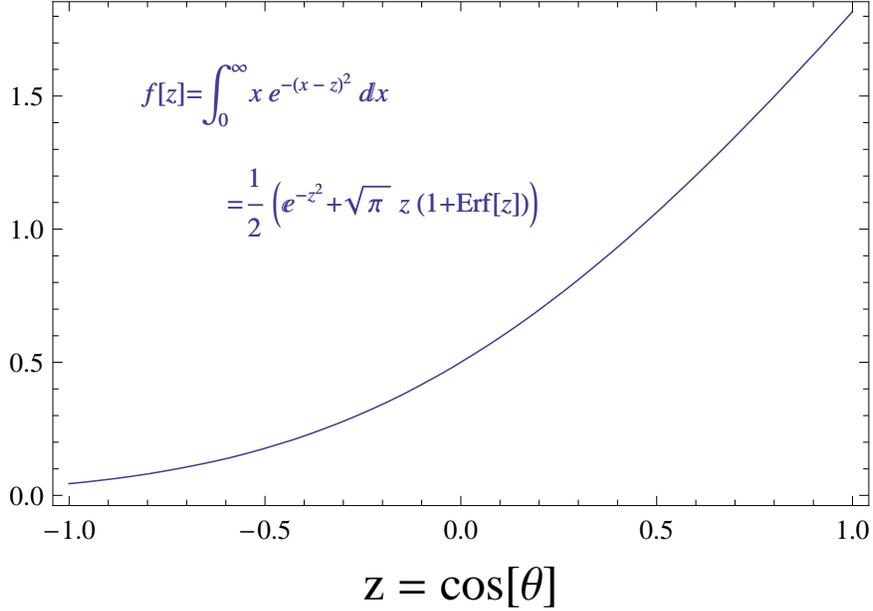}
\caption{\label{Fig1} The angular distribution of the recoils.}
\end{figure}

 Still the width of the $\theta$ distribution together with the high (cubic)  symmetry of the Na-I lattice causes some planar/axial channels to be exposed  to some degree to the recoiling ions at any time. This  dilutes the  dramatic modulations expected if all recoils are directed along the  fixed $W$ directions, but as we indicate below  still leaves significant  daily modulation.

\subsection{III. Possible Experimental Setups}

The latitude of Grand Sasso is $\sim 43^o$ which is close to the $W$ direction. Hence, the WIMP wind hits the top 001 (XY)  plane of the DAMA detectors if indeed placed vertically---almost head-on at the appropriate time of day on both the  wide planar XZ and YZ and the X axis channels are then open.

For this latitude the earth's rotation axis is close to a 101 axis, namely, the face diagonal of the XZ plane.  Hence, unlike the simpler idealized case discussed above the lattice is not invariant under $90^o$ rotation but rather only under a $180^o$. So while the system will have 12-hour  periodicity we may see now even stronger variation after six hours when we will lose all the above channels regaining only the 110 channel which has a smaller acceptance angle!
The scalar product of $\hat{W}$ and $\hat{W}'$---the original---and $90^o$ rotated $W$ directions is
\begin{equation}
 \hat{W} \cdot \hat{W}' = 1/2
\label{scalarprod}
\end{equation}
so that the angle between them is $60^o$. The angular distribution  of incoming WIMPs relative to the direction of $W$'s allows, even at this point in time a nonzero flux of WIMPs to fall on the rotated 100 plane. However the angular distribution of Fig. 1  suggests that this flux is reduced by a factor of 0.5. Thus O(25\%) modulations of the channeling probabilities with 12-hour period are expected. 

 Most existing or planned WIMP direct searches are at $45^o \pm 5^o$ latitude  so that such modulations (with different longitude dependent phases) are expected in all cases.

Can modulations of channeling be translated into specific experimental signatures?

To provide a quantitative answer we need to know how channeling of ions of given energies along specific axes/planes affects the intensities and/or other properties of the e.m. signal in any given experimental setup.

If the channeling effects on enhanced quenching at DAMA's energies are indeed as strong as the analysis of Drubov and DAMA suggest---and as should be the case if channeling is a key in reconciling DAMA and the other
experiments---then also the direction effect and attendant daily variation of scintillations should be seen.

 If the individual events from the recently published DAMA/LIBRA experiment\cite{DAMA1} are appropriately time stamped, and the orientation of the crystals is known, the search for sub-daily modulation of the scintillation signal can be done with existing data! Needless to say, if indeed the expected daily variations with the correct phases will be seen then the likelihood that DAMA has indeed discovered a WIMP dark matter will be greatly enhanced.
Using an old quotation for the DAMA data from which an annual DFS effect has been extracted, ``There could be more gold in them mountains."

 To test the effect of different ion recoil directions the crystal can be exposed to a monochromatic neutron beam (with energy fixed by time of flight, or chopper technique).\cite{Chag},\cite{Sim} From the scattered neutron the energy of the recoiling ion and its direction can be found and the procedure can be repeated for different crystal orientations.  In thin target neutrons, like the WIMPs, collide once in the bulk. By measuring the e.m. response---{\it e.g.,} the scintillation signal in DAMA's NaI(Tl) crystal---in each event some calibration can be directly obtained. Such calibration is indeed precise if the ratios of neutrons' cross sections on the two ions are the same as for the WIMPs.

 Measurements by directly bombarding with the relevant ions with the
 O(keV) energy are susceptible to strong surface effects since such  ions stop in a thin $< 10^3$ Angstrom surface layer.

 We turn next to another class of detectors based on electron hole pair creation for (e.m.) experimental signature.
In high purity germanium detectors the quenching was measured in the first dark matter search utilizing Ge to 
be $\sim$ 25\%\cite{1stDMsearch}(the quenching factor here is defined as the ratio of charge collected in a nuclear recoil event to that induced by a photon of the same energy $E_{recoil}$).
Usually the crystals for the high purity germanium detectors are pulled along the 100 axis by the Czochralski growth technique.  However the remaining 110 and 111 axes can be found by X-ray diffraction or by a technique of the XIA corporation\cite{XIA}  and the Julich group\cite{Jul}.
In the experiments done at the X-ray Instrumentation Associate Laboratory, a weakly collimated $^{57}$Co was used to irradiate the detector with 122 keV gamma rays.  The source was rotated around the vertical 001 axis and the rise time of the pulses was meanwhile continuously measured. Maximal and minimal rise times were recorder about $50^o$ apart. The difference between the maximal and minimal rise times was approximately 20 nanoseconds, whereas the mean rise time was 190 ns. Considering the early finding of Bredov and Okuneva,\cite{Bred}  channeling of the ions may strongly effect the pulse rise times and pulse amplitudes as well.

The genesis of this effect is in the directional dependence of the carrier mobilities.
In a normal p-type closed end coaxial detector, the electric fields point radially inward.
Accordingly an ion liberated by a WIMP and channeled along 110 or 111 axis ({\it i.e.,} not parallel to the cylinder axis) will travel further before stopping conceivably changing significantly the rise time which will thus be sensitive to the initial direction of the WIMP.

Clearly in order to exploit this phenomenon in future Gerda and Majorana arrays should use one of the above X-ray or other techniques to determine the 110 and 111 direction of each individual crystal relative to its mounting. It would also be very valuable to perform the neutron scattering experiments described above and compare the pulse height and rise times from germanium Ions scatters in channels and the rest.
References \cite{XIA},{Jul} provide only information on the directional dependence of the carrier mobility which is related to, but may be quite different from, the directional effects of the WIMP scattering---though channeling effect for WIMPs may well be siginificant.


 \subsection{IV. Summary and Conclusions}

 While the present work was largely motivated by  DAMA's annually varying signal and the potentially crucial role of Ion channeling therein, its  scope is much wider. We focus on the strong directional dependence of channeling and its consequent variation as the crystal which is fixed in a terrestrial laboratory rotates relative to the fixed WIMP Wind daily.
 This in turn will modulate daily the e.m. signal since channeling appears to strongly affect the output e.m. signals from the detectors.
It does so by modifying quenching of scintillation ionic crystal detectors, changing the rise time of the ionization signal in Germanium and potentially in many other ways in other setups. This, in turn, can lead to observed daily modulations in underground detectors due to the WIMP signal

 The purely kinematic DFS annual modulation is expected {\it universally} in all detectors and for all recoil energies. This is not the case for our suggested daily variations which require crystalline detectors and substantial channeling. Still our effect may complement the DFS modulations.

 It is therefore highly worthwhile to look for such a signal in all crystalline detectors and, in particular, in DAMA where an annual signal has been seen and using the very same set of data for yet an independent WIMP test is naturally called for.

 We conclude by listing various, rather obvious points and finally making a few rather speculative additional remarks.
\begin{itemize}
\item[(i)] We focused here on low energy recoils expected for low WIMP masses.
 The more conventional LSP WIMPs have relatively high masses $> \sim$ 100 Gev,
 and for detector nuclei of similar mass we expect high, $\sim$ 10-100 keV,  generic recoil energies.

 This diminishes our effect in two ways.  First, channeling is reduced at  the higher energies. This may be mitigated to some extent by the FF effects  strongly damping the high recoils which $q << 2\mu W$ implies. However, in this
 case the x parameter in Eq. (\ref{angdis}) 
above approaches zero and there is {\it no}
peaking at $\theta=0$ {\it i.e.,} no tendency of the recoils to be in the direction of wind $W$---further undermining our arguments.

\item[(ii)]  The high symmetry of the cubic Na-I crystal helps overall channeling by providing many channeling axes/planes. It does however limit the magnitude of the modulation, suggested as we have several open channels in the angularly broad beam of incident WIMPs at all times.
 One could conceive of using just for this purpose lower symmetry lattices such as  graphite with a pronounced preferred single channeling plane.
Unfortunately most low symmetry crystals consist of ``bad" detector materials and,  in particular, are not transparent to light.

\item[(iii)] Already some time ago, P. Sikivie suggested a complex filament-like pattern for the phase-space of cold dark matter.
Recent large N body simulations and observations of Sagitarius  
flows suggest a complex hierarchial pattern of WIMP density and velocity distribution and it is possible that the local WIMP velocities differ somewhat from the standard scenario underlying the DFS suggestion and the putative DAMA effect. Still, it is most likely that the bulk of the DM is not subclustered and while some modification will occur the basic pattern of WIMP velocities underlying DFS and our suggestion is largely maintained.


\item[(iv)] The dopant Thalium ions tend to scatter the channeling ions in the Na-I crystal and diminish channeling. The latter would thus be much more prevalent  in a crystal of high purity.
 The following ``Wiggler" mechanism  may still allow some e.m. output in this case. The Ion channeling with a speed of $\sim$  100-300 km/sec encounters crystal ions in the wall every $\sim$ a $\sim$ 2-3 Angstrom. This in turn subjects it to a time varying electric  field of frequency $ f=v/a \sim 10^{15}$ sec$^{-1}$ equivalent to 4 eV. If the channeling path is long enough, we have some laser-like coherence and we will be able to keep exciting and de-exciting by spontaneous emission. (Though because of atomic energy scales, this holds only for the hallide ion which here is quite heavy.)

\item[(v)] It is conceivable that channeling will reflect not only in the e.m. signal but in bolometry as well. The recoil speed $\sim$ 100-300 km/sec exceeds by $\sim$ 100 that of sound in the material. We will have therefore a Cherenkov-like cone of supersonic shock with a small opening angle, $\sim c/V \sim$ 1/100. The sound and heat will then arrive first at one particular crystal face, namely, the one facing the favored channeling directions at this time.
\end{itemize}

\subsection{Acknowledgements}
One of the authors (SN) would like to acknowledge helpful discussions with Michele Papucci, Andrew Smith, and Andre Walker-Loud.

\end{document}